\documentclass[a4paper,12pt]{article}
\usepackage[dvips]{graphicx}
\usepackage[dvips]{color}
\usepackage{subfigure}
\usepackage[english]{babel}
\usepackage{amsmath,amsfonts,amsthm,amssymb}

 \newtheorem{thm}{Theorem}[section]

 \newtheorem{prop}[thm]{Proposition}
 \newtheorem{assumption}{Assumption}[subsection]
\newtheorem*{Condition 1}{Condition 1}
\newtheorem*{Condition 2}{Condition 2}

\theoremstyle{definition}
\newtheorem{defn}[thm]{Definition}

\newtheorem{notation}[thm]{Notation}
\newtheorem{defnotation}[thm]{Definition-Notation}

 \theoremstyle{remark}
 
\DeclareMathSymbol{\leqslant}    {\mathrel}{AMSa}{"36}
\DeclareMathSymbol{\geqslant}    {\mathrel}{AMSa}{"3E}

\title{Voronoi Diagram: The Generator Recognition Problem}

\author{M.~Montserrat Alonso Ferrero \\
Dipartimento di Matematica,\\ Universit\'a di Bologna,\\ Piazza di Porta S.~Donato 5,\\ I-40126 Bologna, Italy\\
              E-mail: alonso@dm.unibo.it}
\begin{document}

\renewcommand{\today}{June 30, 2009}

\maketitle

\begin{abstract}

For the analysis of systems consisting of small, regular objects, the methods of mathematical morphology applied to images of these systems are well-suited. One of these methods is the use of Voronoi polygons. It was found that the Voronoi tessellation method represents a powerful tool for the analysis of thin film morphology and provides nanostructural information to many multi-particle assemblies. In these notes, several morphological algorithms are analyzed and we study how to join all of them to design a graphical user interface (GUI) that provides as input for the system the \lq\lq AFM image\rq\rq and interprets the output of the system in terms of errors and generators coordinates.

\end{abstract}

\vskip.3cm



 

\section{Introduction}
In the last decades, there has been an increasing interest in a geometrical construct called the Voronoi diagram (e.g. [1], [2], [3] and [4]).  The Voronoi diagram is a data structure extensively investigated in the domain of computational geometry (e.g. [5]).

Given some number of points in the plane, their Voronoi diagram divides the plane according to the nearest-neighbor rule: Each point is associated with the region of the plane closest to it, so it is a tessellation of $\mathbb{R}^2$. We have already noted that the concept of the Voronoi diagram is used extensively in a variety of disciplines and has independent roots in many of them (e.g. [6]). The first extension of them was to the area of crystallography (the area we are interesting in), works in this field are for example [7] and [8].

\vskip.4cm
Since there is a large number of empirical structures which also involve tessellations of $\mathbb{R}^2$, one of the most direct applications of Voronoi concepts is in the modelling of such structures and the processes that generate them. In these notes, we use the Voronoi assignment model in the modelling of physical-chemical systems. Such systems under study  consist of a set of sites occupied by atoms, ions, molecules, etc. (depending on the specific application) which are represented as equal-size spheres. Our system is formed by sites regularly arranged in $\mathbb{R}^2$, they assume form of lattice (the structure is said to be crystalline). Thin metal films images with Atomic Force Microscopy  (AFM) consist of small two-dimensional islands (objects) distributed on the substrate. The quantitative characterization of the object arrangement can bring information about internal processes in the studied system. We apply methods of mathematical morphology to thin metal films images with Atomic Force Microscopy, to assign the model: The Voronoi Growth Model. Voronoi polygons has been employed for providing nanostructural information to these multi-particle assemblies. We analyze morphological algorithms applied to these tessellations, e.g. to restore the generators from a given Voronoi diagram.

\vskip.4cm
As a graphical user interface (GUI) makes easier for the user to obtain information from algorithms, we present how to join all algorithms, we have studied, to design one. The graphical user interface  provides as input for the system the \lq\lq AFM image\rq\rq, and interprets the output in terms we are interesting in. We note that this work can easily be extended, if we have images from other fields like ecology, meteorology, epidemiology, linguistics, economics, archeology or astronomy, that we suspect are a Voronoi diagram.
\vskip.4cm

The structure of this paper is as follows.
In Section 2 we recall the mathematical theoretical background about Voronoi diagrams and we give an application of them, that it is called the \lq\lq Voronoi Growth Models \rq\rq which we will use in the analysis of film nanographs. In Section 3 we explain the mathematical solution to the problem proposed here. Next in Section 4, we give and analyze algorithms of the mathematical solution and in Section 5 we finish with important concluding remarks and directions for further research.

\section{Preliminaries}

In this section we will review the basic notions we shall require for the sections to follow. For more details about them we refer to [9] and [10] for the first investigation of mathematical aspects of Voronoi diagrams, [11] and [12] for papers that present surveys about Voronoi diagrams and related topics, and [13] for a good introduction to all applications of Voronoi diagrams to sciences.


\subsection{Mathematical Background} \label{seccion 2.2}

We will define the Voronoi diagram and introduce properties and notations to be commonly used in this notes.

\vskip.3cm
We work with a finite number, $n$, of points in the Euclidean plane, and assume that \linebreak $2\leq n<\infty$. The $n$ points are labeled by $p_1, \cdots ,p_n$ with the Cartesian coordinates $(x_{11},x_{12}), \cdots ,\linebreak(x_{n1},x_{n2})$ or location vectors $\textbf{x}_1,\cdots, \textbf{x}_n$. The $n$ points are distinct in the sense that $\textbf{x}_i\neq \textbf{x}_j$ for $i\neq j$, $i,j\in I_n=\left\lbrace 1, \cdots , n\right\rbrace$. Let $p$ be an arbitrary point in the Euclidean plane with coordinates $(x_1,x_2)$ or location vector $\textbf{x}$. Then the Euclidean distance from $p$ to $p_i$ is given by $$d(p,p_i)=\parallel\textbf{x}-\textbf{x}_i\parallel=\sqrt{ (x_1-x_{i1})^2+(x_2-x_{i2})^2}.$$ If $p_i$ is the nearest point from $p$ or $p_i$ is one of the nearest points from $p$, we have the relation $\parallel\textbf{x}-\textbf{x}_i\parallel\leq \parallel\textbf{x}-\textbf{x}_j\parallel$
 for $j\neq i$, $j\in I_n$. In this case, $p$ is assigned to $p_i$. Therefore,

\begin{defn}
Let $P=\left\lbrace p_1, \cdots,p_n \right\rbrace $ where $2\leq n< \infty$ and $\textbf{x}_i\neq \textbf{x}_j$ for $i\neq j$, $i,j\in I_n$. We call the region given by \begin{align} \label{eqdefVor}
V(p_i)=\left\lbrace \textbf{x} \mbox{~s.t.~} \parallel\textbf{x}-\textbf{x}_i\parallel\leq \parallel\textbf{x}-\textbf{x}_j\parallel \mbox{~for~}j\neq i, j\in I_n\right\rbrace 
\end{align}
the \emph{(ordinary) Voronoi polygon} associated with $p_i$ (or the Voronoi polygon of $p_i$), and the set given by 
$$\mathcal{V}=\left\lbrace V(p_1), \cdots, V(p_n)\right\rbrace $$
the \emph{(planar ordinary) Voronoi diagram} generated by $P$ (or Voronoi diagram of $P$).
\end{defn}

We can extend the above definition to the $m$-dimensional Euclidean space, but for our proposes we only need the Euclidean plane. So, we shall often refer to a planar ordinary Voronoi diagram simply as a \emph{Voronoi diagram} and an ordinary Voronoi polygon as a Voronoi polygon.

\vskip.5cm

For a Voronoi diagram $\mathcal{V}$ we have the following definitions.
\begin{defn}
We call the $p_i$ of $V(p_i)$ the \emph{generator point} or \emph{generator} of the $i$th Voronoi polygon, and the set $P=\left\lbrace p_1, \cdots,p_n \right\rbrace $ the \emph{generator set} of the Voronoi diagram $\mathcal{V}$ (Figure 1).
\end{defn}

\begin{notation}
For brevity we may write $V_i$ for $V(p_i)$. Also we may use $V(x_{i1},x_{i2})$ or $V(\textbf{x}_{i})$ when we want to emphasize the coordinates or location vector of the generator $p_i$. In addition, we may use $\mathcal{V}(P)$ when we want to explicitly indicate the generator set $P$ of $\mathcal{V}$.
\end{notation}

\begin{figure}[htbp]
 \centering
 \begin{minipage}[c]{.45\textwidth}
    {\includegraphics[width=6cm]{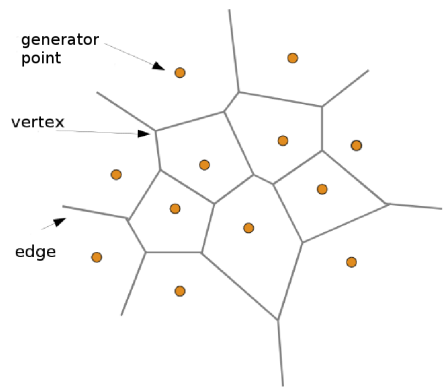}}
    \caption{Voronoi diagram}
  \end{minipage}%
 
\end{figure}
\begin{defnotation}
Given a Voronoi diagram $\mathcal{V}$, since a Voronoi polygon is a closed set, it contains its boundary denoted by $\partial V(p_i)$. The boundary of a Voronoi polygon may consist of line segments, half lines or infinite lines, which we call \emph{Voronoi edges}. Noticing that $=$ is included in the relation of equation (\ref{eqdefVor}), we may alternatively define a Voronoi edge as a line segment, a half line or an infinite line shared by two Voronoi polygons with its end points. Mathematically, if $V(p_i)\cap Vp_j)\neq\emptyset$, the set  $V(p_i)\cap Vp_j)$ gives a Voronoi edge (which may be degenerate into a point). We use $e(p_i,p_j)$ for $V(p_i)\cap Vp_j)$, which is read as the Voronoi edge generated by $p_i$ and $p_j$. Note that $e(p_i,p_j)$ may be empty. If $e(p_i,p_j)$ is neither empty nor a point, we say that the Voronoi polygons $V(p_i)$ and $V(p_j)$ are \emph{adjacent}. 

An end point of a Voronoi edge is called a \emph{Voronoi vertex}. Alternatively, a Voronoi vertex may be defined as a point shared by three or more Voronoi polygons. We denote a Voronoi vertex by $q_i$ (see Figure 1). When there exits at least one Voronoi vertex at which four or more Voronoi edges meet in the Voronoi diagram $\mathcal{V}$, we say that $\mathcal{V}$ is \emph{degenerate} (Figure 2); otherwise, we say that $\mathcal{V}$ is \emph{non-degenerate}. 
\end{defnotation}

\begin{figure}[htbp]
    \centering
    {\includegraphics[width=6cm]{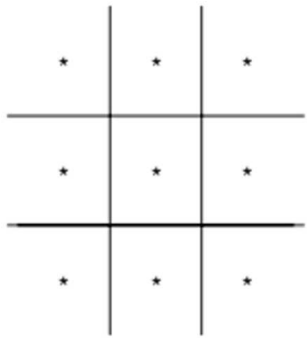}}
    \caption{Degenerate Voronoi diagram}
\end{figure}

In the previous definitions of Voronoi diagram $\mathcal{V}$, we have defined a Voronoi diagram in an unbounded plane. In practical applications, however, we often deal with a bounded region $S$, where generators are placed. In this case we consider the set given by
$$\mathcal{V}\cap S=\left\lbrace V(p_1)\cap S,\cdots ,V(p_n)\cap S\right\rbrace .$$

We observed that an ordinary Voronoi diagram consists of polygons, as a polygon can be defined in terms of half planes, we have the  equality of Proposition \ref{perpendicular2}. 

\begin{notation}
Given a Voronoi diagram $\mathcal{V}(p_1, \cdots , p_n)$, we consider the line perpendicularly bisecting the line segment $\overline{p_i p_j}$ joining two generators $p_i$ and $p_j$. We call this line the \emph{bisector} between $p_i$ and $p_j$ and denote it by $b(p_i,p_j)$. 

Since a point on the bisector $b(p_i,p_j)$ is equally distant from the generators $p_i$ and $p_j$, $b(p_i,p_j)$ is written as
\begin{align*}
 b(p_i,p_j)=\left\lbrace \textbf{x}\mbox{~s.t.~} \parallel
\textbf{x}-\textbf{x}_i\parallel=\parallel\textbf{x}-\textbf{x}_j\parallel\right\rbrace,\mbox{~~} j\neq i.
\end{align*}
The bisector divides the plane into two half planes and gives
\begin{align*}
H(p_i,p_j)=\left\lbrace \textbf{x}\mbox{~s.t.~} \parallel \textbf{x}-\textbf{x}_i\parallel\leq\parallel\textbf{x}-\textbf{x}_j\parallel\right\rbrace,\mbox{~~} j\neq i.
\end{align*}
We call $H(p_i,p_j)$ the \emph{dominance region of} $p_i$ over $p_j$.
\end{notation}

\begin{prop}\label{perpendicular2}
Let $P=\left\lbrace p_1,\cdots , p_n\right\rbrace \subset \mathbb{R}^2$, where $2\leq n<\infty$ and $\textbf{x}_i\neq \textbf{x}_j$ for $i\neq j$, and $i,j\in I_n$. Then
\begin{align*}
 V(p_i)=\bigcap_{j\in I_n\setminus\left\lbrace i\right\rbrace } H(p_i,p_j)
\end{align*}
where $V(p_i)$ is the \emph{(ordinary) Voronoi polygon} associated with $p_i$ and set
\begin{align*}
 \mathcal{V}(P)=\left\lbrace V(p_1), \cdots , V(p_n)\right\rbrace 
\end{align*}
where $\mathcal{V}(P)$ is the \emph{(planar ordinary) Voronoi diagram} generated by $P$.

\end{prop}

As a degenerate Voronoi diagram requires special lengthy treatments which are not essential we avoid this difficulty and we often make the following assumption:

\begin{assumption} \label{asumir1}
(\textbf{the non-degeneracy assumption}). Every Voronoi vertex in a Voronoi diagram is incident to exactly three Voronoi edges.
\end{assumption}.

\noindent \textbf{The largest empty circle in a Voronoi diagram.}
\vskip.4cm
\begin{defn}
For a given set $P$ of points, if a circle does not contain any points of $P$ in its interior, the circle is called an \emph{empty circle}.
\end{defn}

\begin{thm} \label{VDcircle}
Let $Q=\left\lbrace q_1,\cdots ,q_{n_{\mathcal{V}}} \right\rbrace $ be the set of Voronoi vertices of a Voronoi diagram generated by $P$. For every Voronoi vertex, $q_i\in Q$, there exists a unique empty circle $C_i$ centered at $q_i$ which passes through three or more generators. Under the non-degeneracy assumption, $C_i$ passes through exactly three generators (Figure 3).
\end{thm}



From this theorem, the non-degeneracy assumption (Assumption \ref{asumir1}) is equivalent to the following assumption.

\begin{assumption} 
(\textbf{the non co-circularity assumption}) Given a set of points \linebreak $P=\left\lbrace p_1, \cdots ,p_n \right\rbrace \subset \mathbb{R}^2$ ($4\leq n<\infty$), there does not exist a circle, $C$, such that $p_{i1}, \cdots , p_{ik}$, $k\geq 4$, are on $C$, and all points in $P\setminus \left\lbrace p_{i1}, \cdots , p_{ik} \right\rbrace $ are outside $C$.
\end{assumption}

\begin{figure}[htbp]
    \centering
    {\includegraphics[width=6cm]{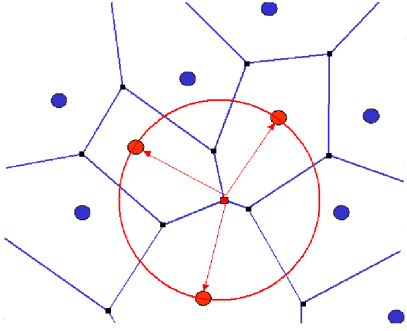}}
    \caption{Empty circle centered at a vertex of a Voronoi diagram}
\end{figure}

\begin{thm}
Circle $C_i$ in Theorem \ref{VDcircle} is the largest empty circle among empty circles centered at the Voronoi vertex $q_i$.
\end{thm}

\vskip.5cm
Here we have seen some properties of the Voronoi diagram, but it has many more. For example, if one connects all the pairs of sites whose Voronoi cells are adjacent then the resulting set of segments forms a triangulation of the point set, called the \emph{Delaunay triangulation}.

\subsection{Voronoi Growth Models}

Thin metal films deposited on a surface consist in their initial stage of growth of small islands. Basic information about nucleation processes during the thin film growth can be derived by the morphological  analysis of the film AFM image. For very thin metal film or generally for systems consisting of small regular objects, the methods of mathematical morphology are well-suited to the study of spatial distribution of objects in images (e.g. [14]). There is a large number of empirical structures which involves tessellations of $\mathbb{R}^2$ (and more generally in $\mathbb{R}^m$), one of the most obvious direct applications of Voronoi concepts is in the modelling of such structures and the processes that generate them. These models produce spatial patterns as the result of a simple growth process with respect to a set of $n$ points (nucleation sites), $P=\left\lbrace p_1,\cdots ,p_n \right\rbrace $, at positions $\textbf{x}_1,\cdots ,\textbf{x}_n$, respectively, in $\mathbb{R}^2$ or a bounded region of $\mathbb{R}^2$. If we make the following assumptions, the resulting pattern will be equivalent to the ordinary Voronoi diagram $\mathcal{V}(P)$ of $P$:

\begin{assumption}\label{ass1}
 Each point $p_i$ ($i=1,\cdots ,n$) is located simultaneously.
\end{assumption}

\begin{assumption}\label{ass2}
 Each point $p_i$ remains fixed at $\textbf{x}_i$ throughout the growth process.
\end{assumption}

\begin{assumption}\label{ass3}
Once $p_i$ is established, growth commences immediately and at the same rate $l_i$ in all directions from $p_i$.
\end{assumption}

\begin{assumption}\label{ass4}
 $l_i$ is the same for all members of $P$.
\end{assumption}

\begin{assumption}\label{ass5}
Growth ceases whenever and wherever the region growing from $p_i$ comes into contact with that growing from $p_j$ ($j\neq i$).
\end{assumption}

Together, Assumptions \ref{ass1}-\ref{ass5} define the \emph{Voronoi Growth Model}. The Figure 4 shows a series in stages in such a growth process.

\hskip .4cm
\begin{figure}[htbp]
 \centering
 \begin{minipage}[c]{.40\textwidth}
  \centering
 \includegraphics[width=4.5cm]{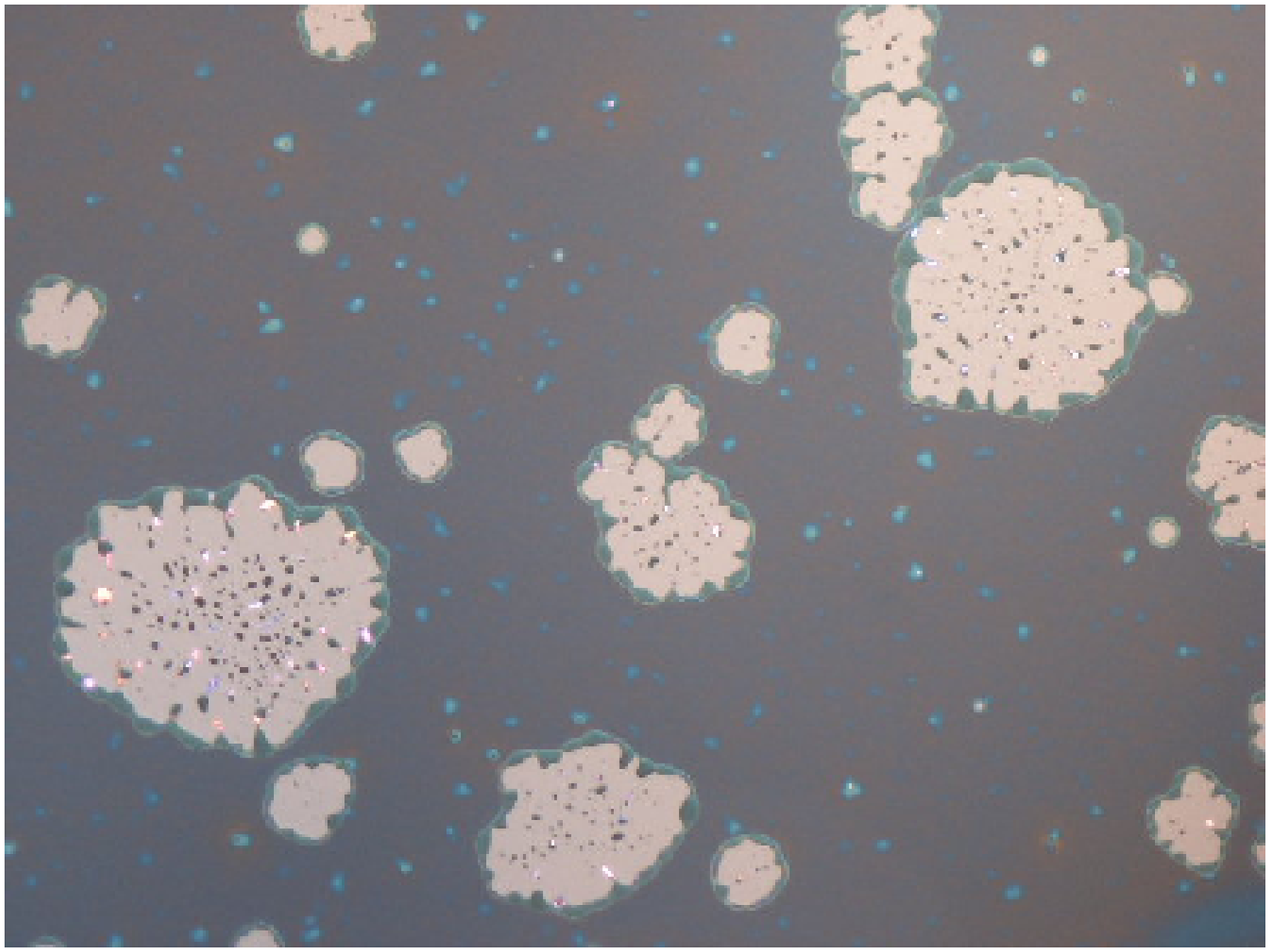}%
 \end{minipage}%
 \hspace{10mm}%
 \begin{minipage}[c]{.40\textwidth}
 \centering
 \includegraphics[width=4.5cm]{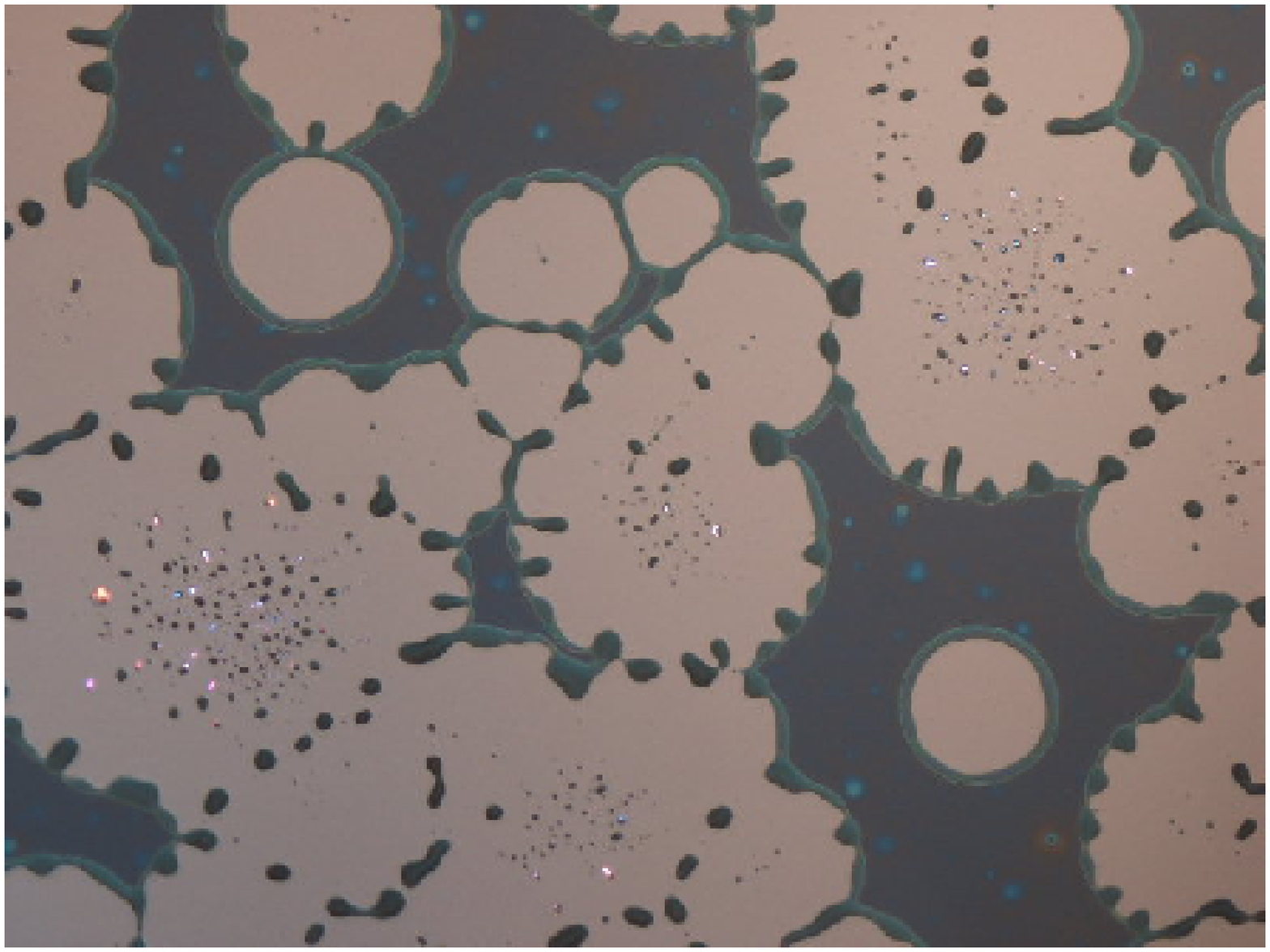}%
 \end{minipage}
\vskip.5cm
\begin{minipage}[c]{.40\textwidth}
   \centering
   \includegraphics[width=4.5cm]{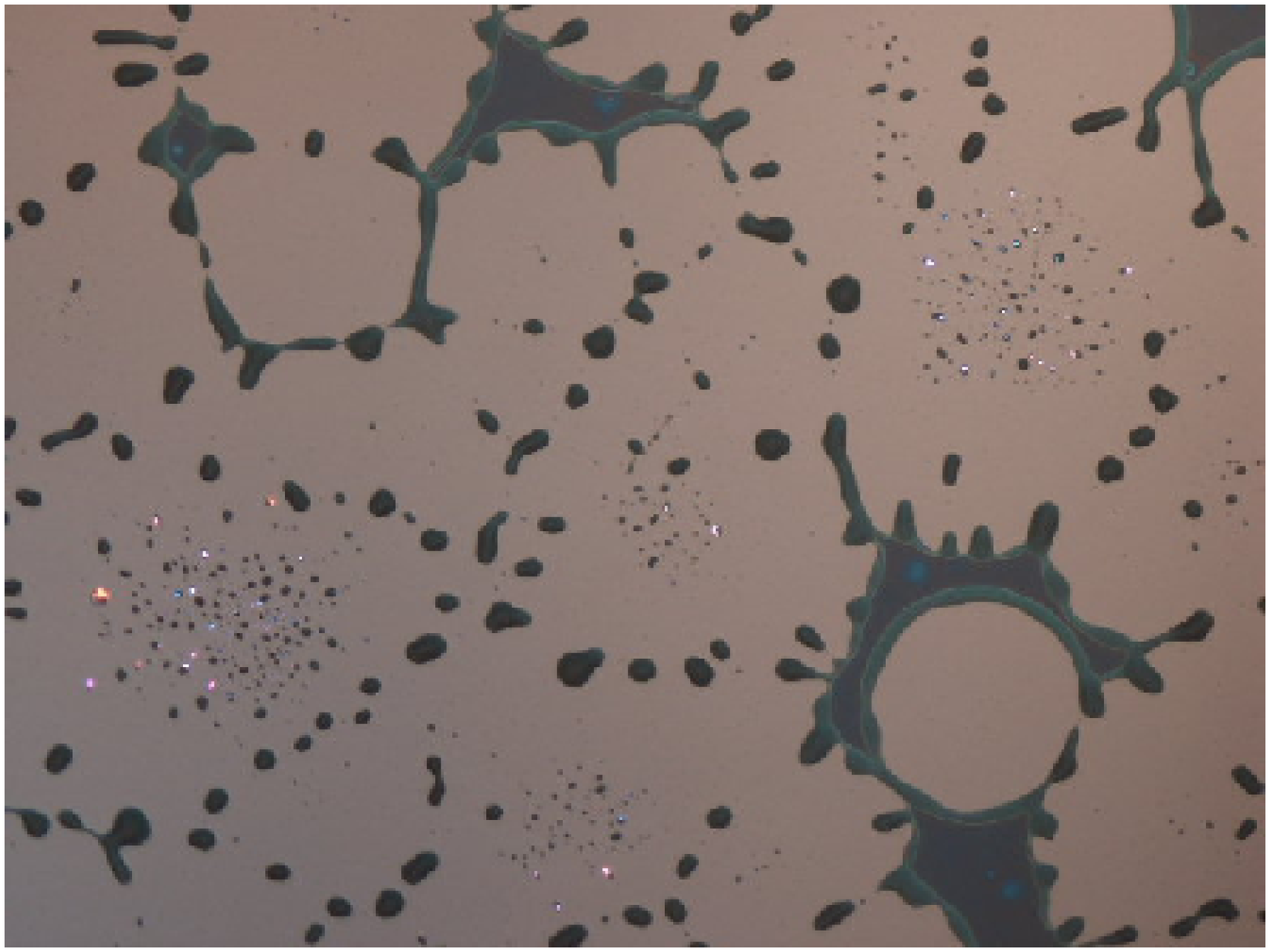}%
\end{minipage}%
\hspace{10mm}%
\begin{minipage}[c]{.40\textwidth}
 \centering
  \includegraphics[width=4.5cm]{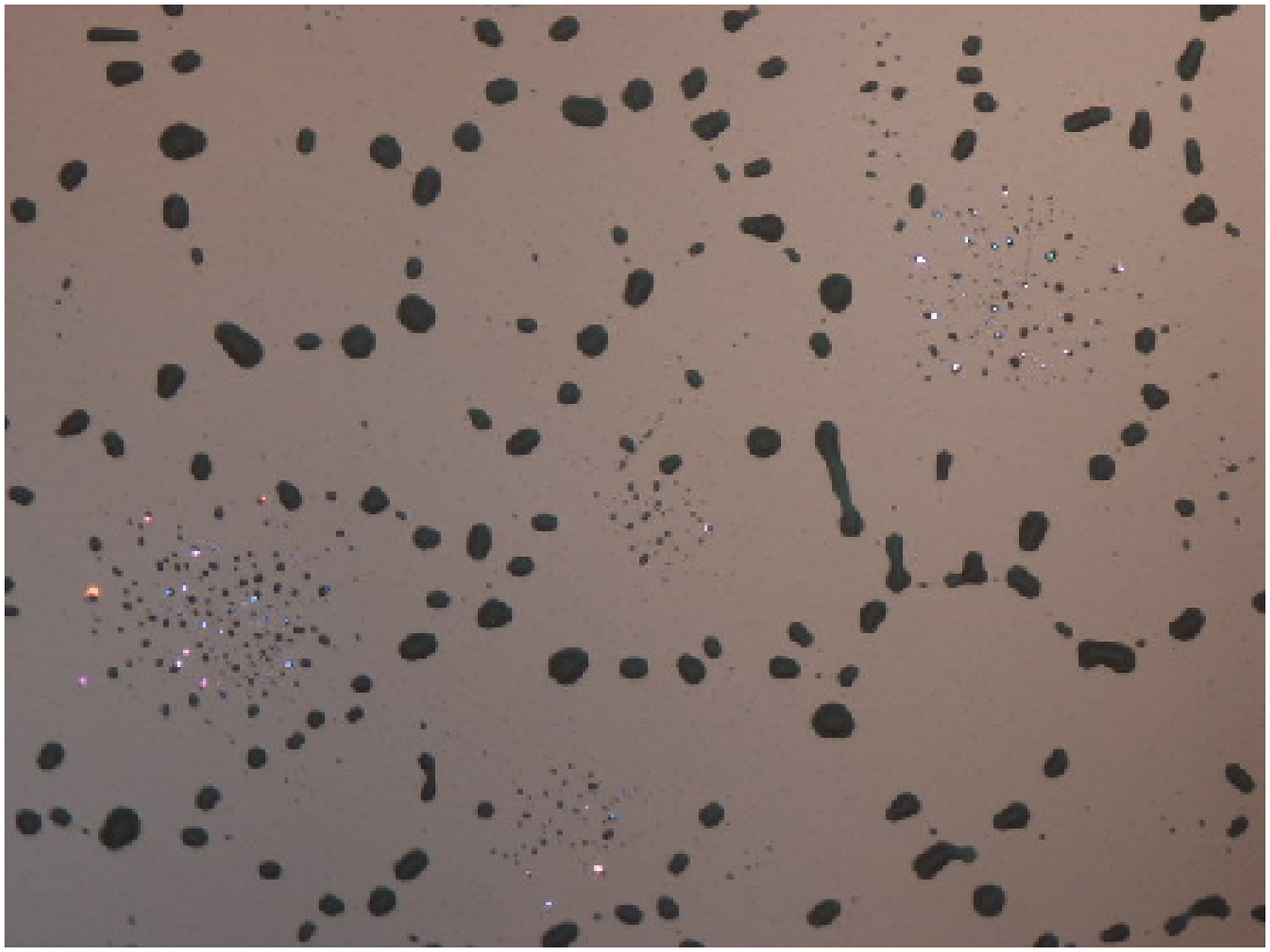}%
\end{minipage}
\caption{Growth process of crystals (Courtesy of Pablo Stoliar)}
\end{figure}

Generally speaking, one obvious application is to model crystal growth about a set of nucleation sites. Here assumptions \ref{ass1}-\ref{ass5} are equivalent to assuming an omni directional, uniform supply of crystallizing material to all faces of the grind crystal in the absence of any absorbable impurities. Assumption \ref{ass3} also implies that the rate of growth of the volume of a crystal will be proportional to its surface area.

Also, growth models to modelling phase transitions in metallurgy involving transformation of an isotropic, one-component solid through nucleation, and isotropic growth of grains of a new or re-crystallized phase. In this context the Voronoi growth model is sometimes referred to as the \emph{cell model} or the \emph{site saturation model}. Specific examples include the covering of a metallic surface by films or layers of corrosion product where the nucleation sites $P$ might be surface imperfections such as impurities, points of intersection with bulk defects and surface pits. Another example is the growth of thin films of metal or semiconductors. In these examples if the thickness of the film is small relative to the spacing between the nucleation sites or if the grain boundaries are perpendicular to the plain of the film, a two-dimensional representation is appropriate.

\section{Problem-Solution}

If the Voronoi assignment and growth models described in the previous section are appropriate for modelling a particular phenomenon, we would expect spacial patterns of the phenomenon to display characteristics of Voronoi diagrams. In  case  we have a tessellation $\mathcal{S}$, we have to  consider ways of determining if it is a Voronoi diagram based on some set $P$ (this problem has been studied e.g. in [15]). Recognizing a Voronoi diagram is closely related with the next generator recognition problem.
\vskip.4cm
\noindent \textbf{The generator recognition problem:} Provided that the Voronoi edges of a non-degenerate Voronoi diagram $\mathcal{V}(P)$ are given, we recover the locations of generators $P$.
\vskip.4cm

\subsection{The generator recognition problem}\label{generadores}
The first problem we approach is to restore the generators from a given Voronoi diagram, that is, the inverse problem of constructing the Voronoi diagram from the given points. For this problem itself, we propose the following geometrical approach (see e.g. [13]).
\vskip.4cm

Let $q_i$ be a Voronoi vertex, $p_{i1},p_{i2},p_{i3}$ be generators whose Voronoi polygons share $q_i$, and $q_{i1},q_{i2},q_{i3}$ be the Voronoi vertices of the Voronoi edges incident to $q_i$. From Theorem \ref{VDcircle}, $q_i$ is the center of the circle that passes through $p_{i1},p_{i2},p_{i3}$. Since Voronoi edges $e(p_{i1},p_{i2})$, $e(p_{i2},p_{i3})$ and $e(p_{i3},p_{i1})$ perpendicularly bisect line segments $\overline{p_{i2}p_{i2}}$, $\overline{p_{i2}p_{i3}}$, $\overline{p_{i3}p_{i1}}$, respectively, we have the equations:

$$\angle p_{i1}q_iq_{i1}=\angle p_{i3}q_iq_{i1}=\alpha _i, $$
$$\angle p_{i1}q_iq_{i2}=\angle p_{i2}q_iq_{i2}=\beta _i, $$
$$\angle p_{i2}q_iq_{i3}=\angle p_{i3}q_iq_{i3}=\gamma _i.$$

Hence $2\alpha_i+2\beta_i+2\gamma_i=2\pi$, i.e. $\alpha_i=\pi-\beta_i-\gamma_i=\pi-\angle q_{i2}q_iq_{i3}$. From this equation, we obtain the following theorem.
\begin{thm}\label{teoremaSolucion}
Let $\overline{q_{i}q_j}$ be a Voronoi edge in a non-degenerate Voronoi diagram, $\theta_{ik}$ and $\theta_{jk}$, $k=1,2,3$ be the acute angles at $q_i$ and $q_j$, respectively, where $k$ is indexed counterclockwise from $\overline{q_{i}q_j}$ at $q_i$ and clockwise at $q_j$. Let $L_{ik}$ ($L_{jk}$) be the half line radiating from $q_i$ ($q_j$) with angle $\pi-\theta_{i2}$ ($\pi-\theta_{j2}$) with $\overline{q_{i}q_j}$ in the sector of $\theta_{ik}$ ($\theta_{jk}$), $k=1,3$ . Then the intersection point made by $L_{i1}$ and $L_{j1}$, and that by $L_{i3}$ and $L_{j3}$ give the generators of the Voronoi diagram sharing $\overline{q_{i}q_j}$.
\end{thm}

We  develop this theorem into a more general theorem with which we can examine whether or not a given planar tessellation $\mathcal{S}=\left\lbrace S_1,\cdots ,S_n\right\rbrace $, is a Voronoi diagram. We suppose that the tessellation $\mathcal{S}$ consists of convex polygons and every vertex has exactly three edges. Let $q_{i1},\cdots ,q_{ik_{i}}$ be the vertices of a polygon $S_{i}$ indexed counterclockwise. Let $p_{ij}$ be the intersection point in $S_{i}$ obtained through the same procedure stated in the previous theorem, where  $\overline{q_{i}q_j}$ is replaced by $\overline{q_{ij}q_{ij+1}}$, $j=1,\cdots, k_{i}$ ($k_i+1$ should be read as $1$). Then we have the main theorem.

\begin{thm}
A planar tessellation consisting of convex polygons whose vertices are all degree three is a Voronoi diagram if and only if $p_{i1}=p_{i2}=\cdots=p_{ik_{i}}$ holds for $i\in I_{n_{v}}$, where $p_{ij}$ is defined in the above.
\end{thm}

\vskip.5cm

An alternative  method to find the generators $P=\left\lbrace p_1,\cdots,p_n\right\rbrace $ of a Voronoi diagram $\mathcal{V}$ proposed in [16] is the following algebraically method. It is based on the perpendicular bisector property i.e. the line segment $\overline{p_{k}p_{l}}$ joining the generators $p_k$ and $p_l$ of two adjacent Voronoi polygons $V(p_k)$ and $V(p_l)$ is bisected perpendicularly by the common edge $e(p_k,p_l)$ of $V(p_k)$ and $V(p_l)$ (see Proposition \ref{perpendicular2}). So, that means that $p_k$ and $p_l$ are subject to the following conditions:
\begin{Condition 1}
$p_k$ and $p_l$ lie on a line perpendicular to $e(p_k,p_l)$.
\end{Condition 1}

\begin{Condition 2}
 $p_k$ and $p_l$ are equidistant from $e(p_k,p_l)$.
\end{Condition 2}

These conditions can be formulated algebraically to form a linear system of equations which can be solved to find the locations $p_k$ and $p_l$. Let $(s_1,s_2)$ and $(t_1,t_2)$ be the locations of the end points of the common edge $e(p_k,p_l)$ of two adjacent members $V_k$ and $V_l$ of $\mathcal{V}$. We search $(x_k,y_k)$ and $(x_l,y_l)$ the locations of the generators $p_k$ and $p_l$, respectively. The segment $e(p_k,p_l)$ lies on the line

$$(s_2-t_2)X-(s_1-t_1)Y+t_2(s_1-t_1)-t_1(s_2-t_2)=0$$ 

\noindent As $\mathcal{V}$ is a Voronoi diagram, Condition 1 gives
\begin{equation*} \label{cond1}
 (x_k-x_l)(s_1-t_1)+(y_k-y_l)(s_2-t_2)=0
\end{equation*}
and Condition 2 gives
\begin{equation*} \label{cond2}
(\dfrac{s_2-t_2}{s_1-t_1})x_k-y_k+(\dfrac{s_2-t_2}{s_1-t_1})x_l-y_l=-2\left( s_2-(\dfrac{s_2-t_2}{s_1-t_1})s_1\right).
\end{equation*}

Suppose that $\mathcal{V}$ is non-degenerate and has $m$ interior edges. Condition 1 gives a system of $m$ equations and $2n$ unknows, and Condition 2 gives another system of $m$ equations and $2n$ unknows. Taken jointly all equations we have enough constraints to provide a least squares solution for $(x_1,y_1,x_2,y_2,\cdots,x_n,y_n)$. Specific methods for solving this equations are given in [16]. The point is that if $\mathcal{V}$ is a Voronoi diagram, then all equations will yield the same solution for $(x_1,y_1,x_2,y_2,\cdots,x_n,y_n)$. Evans and Jones in [16] outline three algorithms for its solution. Unfortunately, the algorithms require the inversion of
poorly-conditioned matrices and may thus be highly unstable.

\subsection{Approximating Voronoi diagrams}

If the exact Voronoi diagram were given, we could determine the position of the generators by the previous methods of Subsection \ref{generadores}. However, such a situation is unrealistic. Recognizing that the recording of many types of empirical patterns often involve some measurement error, it is usual that given a pattern could not correspond to a Voronoi diagram even when we suspect that the pattern was generated by processes such as those in the Voronoi diagram. Even if theoretical consideration tells us that the diagram which appears in a phenomenon should be a Voronoi diagram, the error in observation process must perturb the original diagram. Therefore, the geometrical method would always tell us that the diagram is not the Voronoi, i.e., it would give us no information in almost every case. Methods proposed in [16], [17], [18]  tell us at least approximate positions of the generators.

\section{Voronoi diagram in Computational geometry: Algorithms}

Geometric objects such as points, lines, and polygons are the basis of a broad variety of important applications and give rise to an interesting set of problems and algorithms. Computers are being used to solve larger-scale geometric problems. \emph{Computational Geometry} has been developed as a set of tools and techniques that takes advantage of the structure provided by geometry.

Now, we describe algorithms to solve the generator recognition problem (see [PFL]). First of all we must store the tessellation of which we are seeking the generators. A tessellation is typically stored as a list of vertex coordinates and its associated contiguity lists: lists which provide, for each vertex, the indices of the other vertices to which it is connected. If a vertex lies on an infinite edge, we  store both the vertex and an arbitrary other point $\overline{p}$ on the infinite edge, where $\overline{p}$ is labeled a dummy vertex and given no adjacency list. In the input of the algorithms we will describe, we require that the number of ordinary vertices and the number of dummy vertices be specified, and that the dummy vertices be placed at the end of the vertex list (we will consider them as degenerate).

Let $\mathcal{V}$ be a tessellation of the Euclidean plane and $Q=\left\lbrace q_1, \cdots ,q_{n_{v}} \right\rbrace $ the set of vertices  in which the last $n_{c}$ vertices lies in a infinite edge. By Theorem \ref{teoremaSolucion}, if the tessellation is a Voronoi diagram for each vertex $q_j$ of a given polygon $V_i$ we can define a half line $L_{ij}$ of a giving direction radiating from $q_j$ into the interior of $V_i$. The intersection of any two such half lines gives the location of the generator of $V_i$.

[19] gives the implementation of the following algorithms. 

\subsection*{Algorithm I (naive)} 
The above introduction suggests the following naive algorithm for tessellations such that all of whose polygons contain at least two non degenerate vertices:

\vskip.5cm
\noindent ALGORITHM I

\begin{description}
\item[Step 1.] Specify the polygons $V_1, \cdots, V_n$. 
\item[Step 2.] For each polygon $V_i$:
\begin{description}
\item[2.1] Find any two non degenerate vertices outlining $V_i$, say $q_{i1}, q_{i2}$.
\item[2.2] For each vertex $q_{ik}$ ($k=1,2$) find the ray extending from $q_{ik}$ ($k=1,2$) through the generator in $V_i$, as we described above.
\item[2.3] Find the intersection of this two rays.
 \end{description}
\end{description}

\vskip.5cm
\noindent PROBLEMS

\begin{description}
\item[i)] The requirement that each cell contain at least two non degenerate vertices. 
\item[ii)] Only two rays are used to determine the generator in each polygon.
\item[iii)] If the two rays in a polygon are perfectly parallel (a simple modification is to find an additional ray emanating from a different non degenerate vertex in the polygon). 
\end{description}

\subsection*{Algorithm II} 

Errors  in the generator determination of the previous algorithm could be minimized by using all the available rays rather than just two. Hence an alternative is the following algorithm:

\vskip.5cm
\noindent ALGORITHM II

\begin{description}
\item[Step 1.] Specify the polygons $V_1, \cdots, V_n$. 
\item[Step 2.] For each polygon $V_i$:
\begin{description}
\item[2.1] Find all non degenerate vertices outlining $V_i$.
\item[2.2] For each vertex find the ray associated.
\item[2.3] Find the intersection of every possible pair of rays.
\item[2.4] Average these intersection points.
 \end{description}
\end{description}

\vskip.5cm
\noindent PROBLEMS

\begin{description}
\item[i)] The generator location errors using the Algorithm II are in fact typically considerably larger than for Algorithm I!!! 
\end{description}

\subsection*{Algorithm III} 

This increase in error in the previous algorithm is attributable to the instability in intersecting certain select pairs of rays, one may modify Step 2.4 of Algorithm II by computing a weighted average of the intersection points, weighting each point according to an estimate of its stability, as in the following algorithm:

\vskip.5cm
\noindent ALGORITHM III

\begin{description}
\item[Step 1.] Specify the polygons $V_1, \cdots, V_n$. 
\item[Step 2.] For each polygon $V_i$:
\begin{description}
\item[2.1] Find all non degenerate vertices outlining $V_i$.
\item[2.2] For each vertex find the ray associated.
\item[2.3] Find the intersection $p_{lk}$ of every possible pair of rays in the polygon $V_i$.
\item[2.4] For each pair $(k,l)$ of rays in the polygon $V_i$, estimate the stability of its intersection by perturbing the slopes of each of the rays by a small amount in either direction and seeing how much the intersection point changes. Record $\delta_{k,l}$ = the sum of the sizes of these changes.
\item[2.5] Compute a weighted average of the intersection points, giving $p_{lk}$ the weight\\ $(\delta_{k,l})^{-1}/(\Sigma_{k',l'} (\delta_{k',l'})^{-1})$.
\end{description}
\end{description}

\vskip.5cm
Note that a potential alternative to algorithms II and III is to find the point minimizing
some penalty function such as the sum of squared perpendicular distances to the rays. The
(weighted) averaging in algorithms II and III is equivalent to finding the location minimizing the (weighted) sum of squared distances to the intersection points of the rays.

\subsection*{Remarks} 

Algorithms I, II, and III are all entirely local; each polygon is determined solely based on its own vertices and their neighboring vertices. The accuracy of the algorithms can potentially be improved by incorporating information from neighboring polygons, e.g. by using the perpendicular bisector relation of Proposition \ref{perpendicular2}.  Paik, Ferguson and Li suggest modifying Algorithms I, II, and III to improve the results.
\\
All of the algorithms proposed  are extremely fast, requiring just $O(n)$ observations, where $n$ represents the number of generators to be determined. 
\\
The errors in the inversion algorithms proposed  are very small. However, in [19] they inquire about the size of the errors resulting when one of the vertices is recorded substantially in error.

\section{Summary and proposed extensions} \label{example}

We start with an AFM image, like one in the figures below that represent the growth of crystals with different velocities (Courtesy of Pablo Stoliar).

\begin{figure}[!htbp]
 \centering
 \begin{minipage}[c]{.40\textwidth}
 \includegraphics[width=6cm]{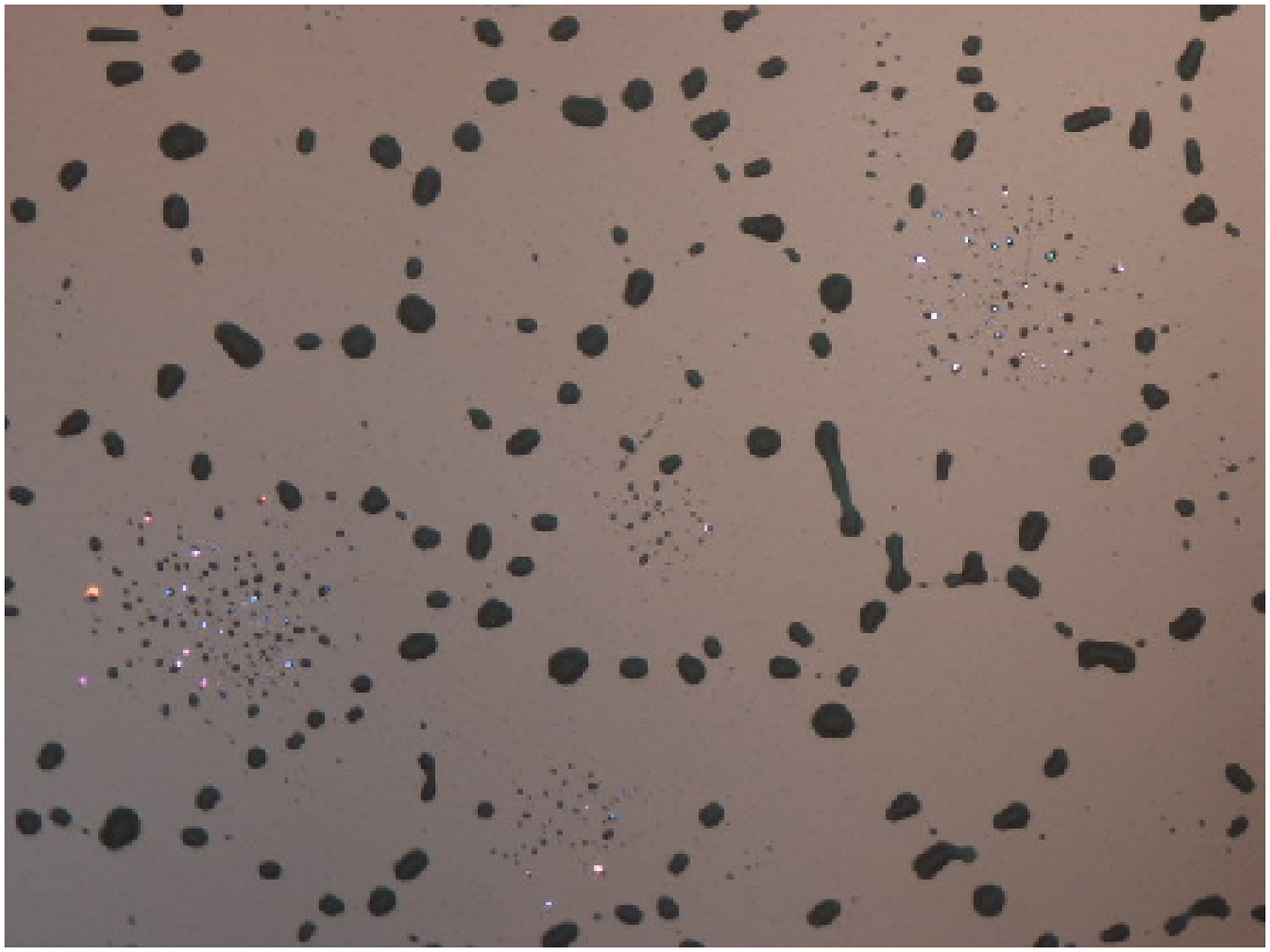}%
 \end{minipage}%
 \hspace{10mm}%
 \begin{minipage}[c]{.40\textwidth}
 \includegraphics[width=6cm]{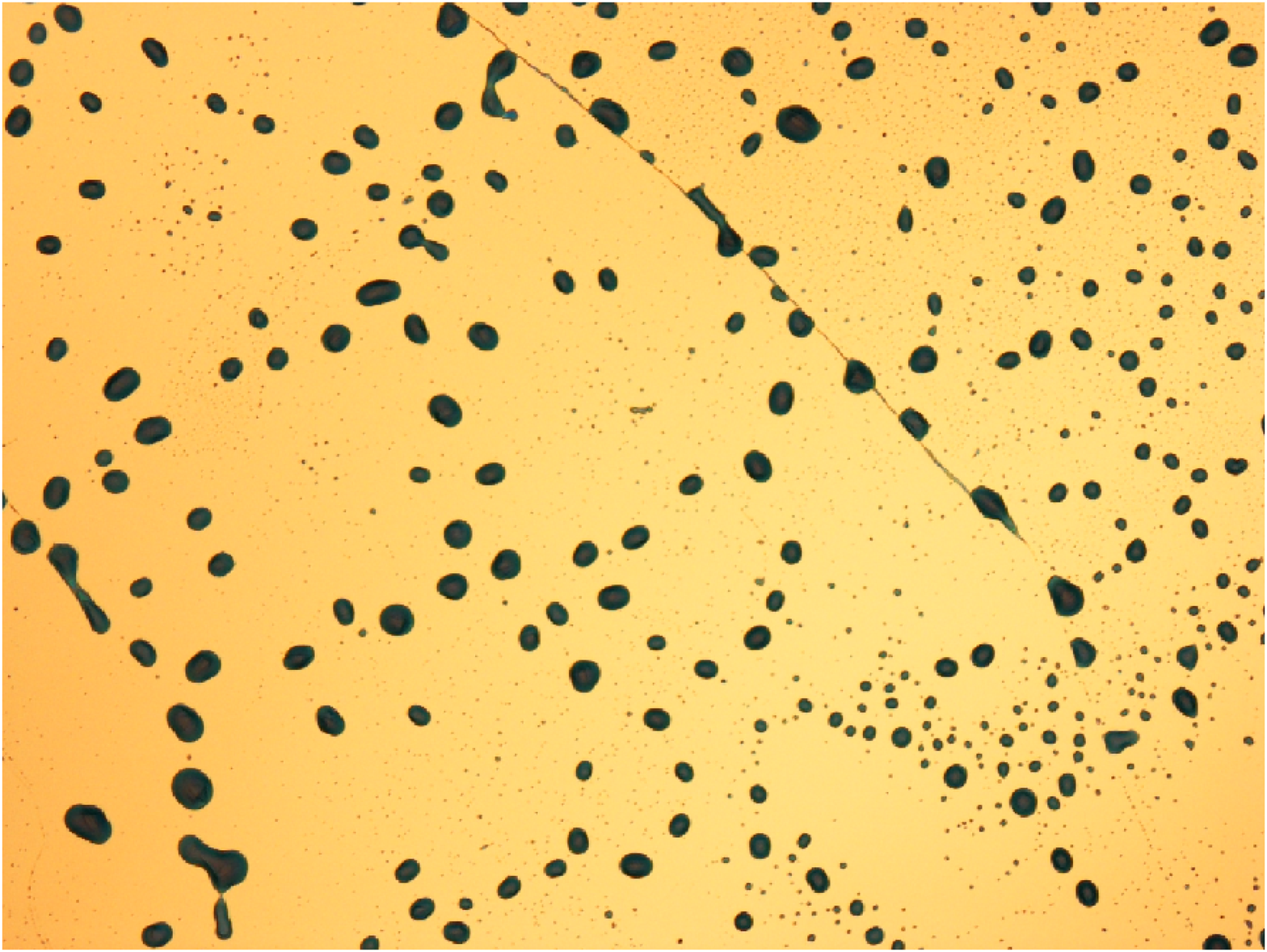}%
 \end{minipage}
\vskip.4cm
 \centering
 \begin{minipage}[c]{.40\textwidth}
 \includegraphics[width=6cm]{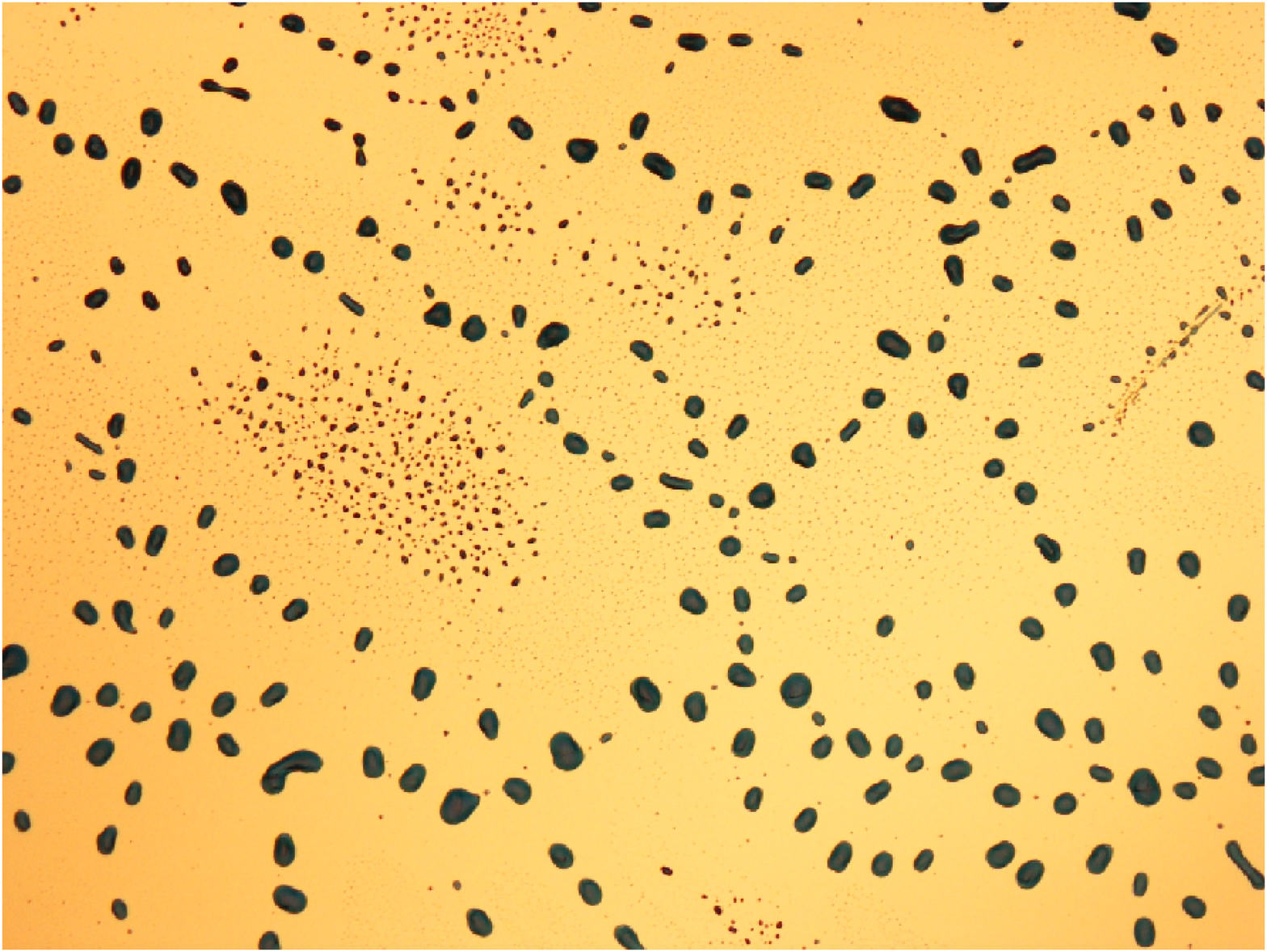}%
 \end{minipage}%
 \hspace{10mm}%
 \begin{minipage}[c]{.40\textwidth}
 \includegraphics[width=6cm]{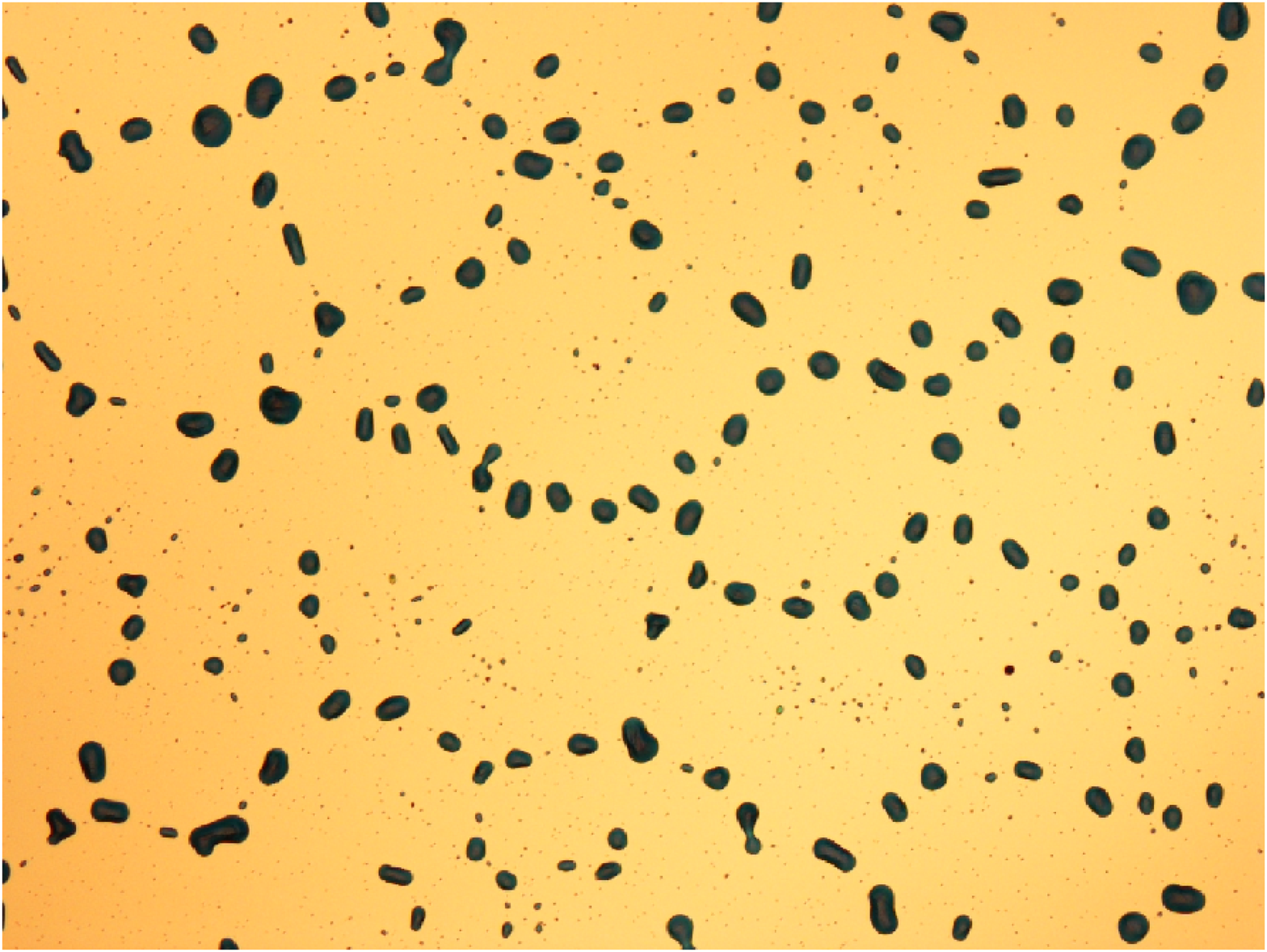}%
 \end{minipage}
\end{figure}

They can be represented as a tessellation of a bounded region of the plane. We want to apply Voronoi diagrams to pattern recognition to this branch of solid state physics. If we suspect that these structures are generated by spatial processes resulting in tessellations which can be constituted by Voronoi diagrams, we would have to follow the next steps to analyze the images in order to obtain properties of the thin films they represent.

\begin{enumerate}
\item  We approximate the image to extract the vertices of the tessellation.
\item  Using one of the algorithms of Subsection 3.2 we approximate this tessellation to a Voronoi diagram.
\item In order to apply algorithms of Section 4, we store the vertices like we said in this section. The algorithms take as input this list of vertices and their adjacency lists, the output give us the generator points of the Voronoi diagram.
\item The last step will be to measure the errors (root-mean-squared errors in the vertices locations) between the tessellation we obtain directly from the image, and the Voronoi diagram we obtain with the outputs of the algorithms I, II and III.   
\end{enumerate}

Finally, the present work may be extended in the following way: every step uses a different computational algorithm, it is interesting to join all this steps to design a graphical user interface (GUI) that provides as input for the system the \lq\lq AFM image\rq\rq, and interprets the output of the system in terms of errors and generators coordinates. A user interface makes easier for the user to interact with the designed programs utilizing toolbar buttons and/or icons. Every software package, the one for step 1, step 2, step 3 and step 4, need a graphical user interface design that can be developed, we think it would be useful to do only one graphical user interface with all these algorithms inside, in that way we can obtain quickly the thin film information we need. 

This graphical user interface could be easily extended to another kind of images, i.e. we can provide as input for the system a variety of images that it could be represented by a Voronoi diagram. So, this work will be valuable not only in the field of crystallography, but also in the fields such as ecology, meteorology, epidemiology, linguistics, economics, archeology and astronomy where Voronoi diagrams are applied.

\end{document}